\documentclass[twocolumn,showpacs,preprintnumbers,prl]{revtex4}
\begin{document}
\title {Two-dimensional Induced Ferromagnetism}
\author { M. K. Mukhopadhyay, M. K. Sanyal}
\affiliation {Surface Physics Division, Saha Institute of Nuclear Physics, 
1/AF, Bidhannagar, Kolkata 700 064, India.}
\author {M. D. Mukadam, S. M. Yusuf}
\affiliation {Solid State Physics Division, Bhabha Atomic Research Centre, 
Mumbai 400085, India.}
\author { J. K. Basu }
\affiliation {Experimental Facilities Division, Advanced Photon Source, 
Argonne National Laboratory, USA}
\date{\today}
\begin{abstract}
Magnetic properties of materials confined to nanometer length scales are 
providing important information regarding low dimensional physics. Using 
gadolinium based Langmuir-Blodgett films, we demonstrate that
two-dimensional ferromagnetic order can be induced by applying
magnetic field along the in-plane (perpendicular to growth) direction.
Field dependent exchange coupling is
evident in the in-plane magnetization data that exhibit absence
of hysteresis loop and show reduction in field required to obtain
saturation in measured moment with decreasing temperature.\\
\end{abstract}
\maketitle

The role of dimensionality in magnetic ordering has remained an active
field of research since the first argument of Bloch in 1930 followed by
theoretical work of Mermin and Wagner in 1966
\cite{bloch,mermin}. Recent advances in growth and characterization
techniques of nanomaterials - materials confined to nanometer length
scales in one, two or all three dimensions - has
enabled investigation of magnetic ordering in general and ferromagnetism
in particular in one-dimensional (1D) \cite{kmnat}, two-dimensional
(2D) \cite{4,5} and zero-dimensional (0D) \cite{6} systems. These
studies have given opportunities to refine existing theoretical
understanding regarding possible existence of long-range
ferromagnetic order in low-dimensional systems \cite{ahar,8,9,10}.
                                               
Pomerantz et al. \cite{10,11}, demonstrated for the first time that
one can form {\em literally} two-dimensional (2D) magnets using
Langmuir-Blodgett (LB) film growth technique \cite{physrep}. Using
LB film growth technique, one can form 2D hexagonal
lattice of metallic ions and multilayer stack of these 2D lattices
can be kept separated by organic chains \cite{physrep,prb02}. In a
typical LB film metal-metal distances within and between the 2D
lattices are about 5 \AA~and 50 \AA~respectively, making this
arrangement ideally suited to form stacks of 2D
lattices of single-ion magnets. Although several magnetic ions like
manganese \cite{lbmag}, iron \cite{lbiron}, cobalt \cite{lbco} have
been used to form LB films, clear signature of 2D ferromagnetic
ordering has not been observed in these systems. Recently, signature
of magnetic ordering was observed in gadolinium based LB films at
unusually high temperature \cite{bohr}. But due to
doubtful stoichiometry and absence of systematic low temperature
magnetic measurements, the nature of ordering could not be established.
The magnetic properties of solids based on gadolinium, a lanthanide
metal, are primarily determined by the localized 4f moments. One can
observe long-range magnetic ordering here provided the exchange
coupling is mediated by the hybridized 6s and 5d conduction
electrons \cite{free, maiti}. On the other hand,
one expects to observe paramagnetism \cite{henry} in gadolinium
compounds due to absence of conduction electrons.
However, in a recent systematic angle resolved photoemission
measurements of oxygen-
induced magnetic surface states of lanthanide metals, it was shown that 
gadolinium forms GdO instead of nonmetallic sesquioxide Gd$_2$O$_3$ 
\cite{schubler}. The remaining one valence electron of ($5d6s^2$)
hybridized state was found to be responsible for mediating exchange
coupling to form magnetic ordering. The temperature dependence of
the energy splitting
of the oxygen-induced states has confirmed \cite{schubler} this
possibility.                                                                  

Here we present evidence of 2D ferromagnetism that can be induced
by external magnetic field applied along the metallic planes in
gadolinium stearate (GdSt) LB film. The gadolinium ions show normal
paramagnetism when the field is applied in the out\,-\,of\,-\,plane
(normal to the 2D gadolinium layers) direction. Several GdSt LB films,
having 9 to 101 monolayers (ML), were deposited on 1 mm
thick Si(001) substrates using an alternating trough (KSV5000) from
a monolayer of stearic acid on Milli-Q (Millipore) water subphase
containing 5$\times 10^{-4}$ M Gd$^{3+}$ ions, obtained from dissolved
gadolinium acetate. The surface pressure was maintained at 30 mNm$^{-1}$
during deposition and the dipping speed was 5 mm\,min$^{-1}$. The
silicon substrates were cleaned and hydrophilized according to RCA
cleaning procedure. Grazing incidence x-ray reflectivity and
diffuse scattering measurements were performed using a rotating anode
x-ray set up, to characterize the structure of deposited LB films
\cite{physrep}.

We have presented magnetization data of 101 ML LB film as a function of
T and H measured using a vibrating sample magnetometer (Oxford
Instruments). The nature of magnetic ordering was found to be
independent of number of monolayers deposited but in 101 ML sample
the ordering was evident in the raw
data itself despite the presence of diamagnetic signal of Si(001)
substrate (refer Fig.1(a)). Magnetization isotherm measurements
(M vs. H at a fixed temperature) over all four quadrants
including the virgin curve were carried out as a function of magnetic
field up to $\pm$70 kOe, applied parallel (in\,-\,plane) as well as
perpendicular (out\,-\,of\,-\,plane) to the film plane at several
temperatures down to 2 K. All these measurements were carried out
by cooling the sample from 300 K to the desired temperature of
measurement under zero magnetic field. No hysteresis was found
in these M vs. H curves. Field-cooled (FC) M vs. T measurements
were carried out over 2 to 100 K under 500 Oe in the cooling cycle.
We also found that the magnetization values scale with number of
monolayers deposited.

Silicon background was subtracted from all the data consistently before 
performing data analysis. Fig.1(b) depicts M vs. T curves measured with
500 Oe field in two in-plane directions, obtained by rotating
the film by 90$^\circ$, and in an out-of-plane direction.
Paramagnetic behavior in
all three directions is evident. However,  at lower temperatures
the nature of the out-of-plane data was found to be different from
that of in-plane data (Fig.1(b)) indicating
different spin response in the in-plane directions as we shall
discuss later. In Fig. 1(c) we have shown the specular x-ray
reflectivity data of 9 ML GdSt LB film. In this data the presence
of both Bragg peaks and Kiessig fringes corresponding to out-of-plane
metal-metal distance and total film thickness, respectively
\cite{gibaud}, are evident. For films with large number of layers,
Bragg peaks become strong and Kiessig fringes could not be resolved
(refer 51 ML data in Fig. 1(c)). The measured 9 ML data
matches quite well with the calculated reflectivity obtained from
a simple model electron density profile shown in the
inset. The organic portion of film has electron density of 0.32
el\,\AA$^{-3}$ and dips going to the value of 0.17 el\,\AA$^{-3}$,
as observed earlier \cite{gibaud}. The electron density of the
head region takes a value of 0.64 el\,\AA$^{-3}$ to
produce the strong Bragg peaks observed in reflectivity data. This
value of electron density in the metal-plane corresponds to a
molecular structure where two stearic acid tails ( having 20
\AA$^2$ area) are attached to a single gadolinium ion \cite{prb02}.
We have also shown the calculated reflectivity data and corresponding
electron density profile in Fig.1(c) assuming that three tails are
attached to single gadolinium ion. The essential difference between
two density profiles being the change of electron density in the
metal-plane from 0.64 el\,\AA$^{-3}$ to 0.48 el\,\AA$^{-3}$. From
these curves, we conclude that out of three valence
electrons in gadolinium only two electrons participate in bonding. 
Diffuse scattering data of these films show clearly
that the 2D metal-planes are conformal in nature and
have logarithmic in-plane correlation, as observed earlier \cite{basu}.
The interfacial roughness comes out to be around 2 \AA.

Figure 2(a) shows out-of-plane magnetization data measured at 5K, 10K
and 20K temperatures. All the magnetization data plotted against (H\,/\,T)
collapses to a single curve as expected for paramagnetism or
superparamagnetism \cite{yusuf}. The data were fitted with the expression
M = M$_s$B$_s$(g$\mu_B$SH/k$_B$T) where
M$_s$ (= Ng$\mu_B$S/V) is the saturation magnetization and B$_s$ is the 
Brillouin function defined as \cite{ahar} 

\begin{equation}
B_s(x)= \frac{2S+1}{2S}coth\left(\frac{(2S+1)x}{2S}\right)
-\frac{1}{2S}coth\left(\frac{x}{2S}\right)
\end{equation}

We obtained the value of spin S as 2.75 instead of expected 3.5 of the
4f moment for gadolinium. M$_s$ (=Ng$\mu_B$S/V) was found to be
1.29 $\times 10^{-5}$emu/mm$^2$. This value corresponds well with
the number density of gadolinium 2.53~$\times 10^{-6}$~mm$^{-2}$. Here,
each gadolinium ion was assumed to be at the center of a cylinder of
20 \AA$^2$  cross section
and length 50 \AA~, as obtained from fitting of the specular
reflectivity data shown in Fig.1(c).

In Fig. 2(b) in-plane magnetization data taken at temperatures of 2K,
5K, 10K and 20K are shown. It is interesting to note that the slope
of the curves as well as the respective saturation magnetization
($M_s$) values decreases as the temperature is increased. As a result,
the in-plane magnetization curves do not
collapse to a single curve like the out-of-plane data ruling out
the existence of normal paramagnetism or superparamagnetism in these
2D planes.

However, like in out-of-plane data, no hysteresis (i.e., zero remanent
magnetization and zero coercive field) was observed here. This type of
field-induced ferromagnetism has been observed earlier \cite{guertin}.
It should be noted here that field values (H$_s$) at which saturation
of magnetization sets in was found to decrease with decreasing
temperature and we obtained values of 10.2, 21.9, 34.6 and 57.2 kOe
for sample temperatures of
2, 5, 10 and 20K, respectively. The saturation magnetization was found
to exhibit 
exponential dependence with temperature ($logM_s=0.66-0.034T$)
(refer inset of Fig.2(b)). Here M$_s$ is expressed in a unit of
$\mu_B$/Gd and the projected value of M$_s$ at 0K comes out to be 4.57 
$\mu_B$/Gd. This value is less than the value of 5.5 obtained from
the fitting of out-of-plane data. It is interesting to note that the
magnetization curves taken in the out-of-plane direction do not show
this saturation behavior, which implies strong anisotropy in GdSt LB
film. In fact, effect of strong anisotropy in ferromagnetism and
possible existence of even direction dependent Curie temperature has
been discussed earlier \cite{26}. We have used an anisotropic exchange
to explain the induced ferromagnetism in the in-plane
direction -- such an exchange was noted to be sufficient to stabilize 
ferromagnetism in 2D systems \cite{mermin}. In this formalism it is expected that 
increase in applied magnetic field in a particular in-plane (xy)
direction will increase the effective exchange field provided
J$_x$,J$_y\gg$J$_z$. The existence of saturation in magnetization
only in the in-plane direction corroborates this assumption.

We could analyze all the in-plane data by using Brillouin function
as used in ferromagnetism \cite{ahar}
\begin{equation}
M=M_sB_s\left(\frac{S}{k_BT}[g\mu_BH + J_{||}
\left<S_j\right>\sum_j cos\theta_{ij}]\right)
\end{equation}
where J$_{||}$=J$_x$=J$_y$ are in-plane exchange (assumed to be symmetric
on the basis of data shown in Fig.1(b)) and $\theta_{ij}$ is the
in\,-\,plane angle between spins S$_i$ and S$_j$.  

With increasing applied field the  {\em sum of cosines}, in Eq. (2),
gets maximized. This  relationship can be used to write a simplified
expression for in-plane magnetization 
\begin{equation}
M = M_s B_s(\frac{SH\mu_B}{k_BT}(g + \xi M))
\end{equation}

In Fig. 3 we have plotted measured M along with the fitted curves
obtained by
using Eq.(3) for 2, 5, 10 and 20K data. In this analysis we have
used S= 3.5 and only fitting parameter was $\xi$, which increases
with increasing temperature. The values come out to be 2.45
$\times 10^{4}$, 8.82 $\times 10^{4}$, 2.15 $\times10^{5}$ and
8.28 $\times 10^{5}$ mm$^2 emu^{-1}$ at 2, 5, 10 and 20K temperatures 
respectively. It should be noted here that without invoking this
field dependent exchange interaction ($\xi$H) the data couldn't be
analyzed. We have tried fitting the data with a Brillouin function
with constant exchange in place of $\xi$H. The best fit was obtained
(refer dashed curve in Fig. 3) with $\xi$H =1.27 $\times 10^{6}$ kOe
mm$^2 emu^{-1}$ for the 5K data. Only field dependent (linear)
exchange could explain the observed magnetization data (refer Fig. 3.).

It will be interesting to investigate the roles of various models of 
ferromagnetism for gadolinium \cite{maiti,schubler} and 2D
dipole interactions \cite{politi} to explain the observed
reduction of H$_s$ and increase of M$_s$ as temperature decreases. Lower 
saturation value of magnetization (M$_s$) in the in-plane direction can be 
simply due to non-participation of few gadolinium planes, located near 
substrate/film and film/air interfaces, in 2D ferromagnetic ordering.
However, strong asymmetry arising from lower spin value (2.75) in the
out-of-plane direction and lower M$_s$ value with expected spin of
3.5 in the in-plane direction need systematic microscopic measurements
to elucidate the effect of anisotropic g-factor for spins \cite{28,
29,halperin}.

In conclusion, we have demonstrated that GdSt LB films can provide an
easy-to-form 2D ferromagnetic system. We have shown that out-of-plane
magnetization
resembles paramagnetic behavior down to a temperature of 2K and a field
of up to 70 kOe. However, field induced ferromagnetic ordering was
observed in the in-plane direction and the field (H$_s$) required to
get the saturation magnetization decreases with decreasing temperature.
The attachment of two stearic acid tails with single gadolinium ion
and the presence of field induced ferromagnetic state in GdSt, found
here, are consistent with the reported results of recent photoemission
studies \cite{schubler}. The absence of remanence in the in-plane
ferromagnetic ordering, observed perhaps for the first time here, provides
experimental validity of Mermin-Wagner theorem which has been extended
recently \cite{bruno} for the metallic two-dimensional systems.

\newpage

\begin{figure}
{\bf Figure captions:}
\caption{(a) Raw data of in-plane magnetization as a function of applied 
magnetic field taken at 5 K for 101 ML GdSt
LB film (line+solid circle) and for bare Si substrate used here (solid line). 
(b) The in-plane magnetization data in two orthogonal directions (line+star)
and (line+triangle) respectively and out-of-plane magnetization data
(line+circle) measured as a function of temperature with an applied field of
0.5 kOe. (c) Open circles are the experimental x-ray reflectivity data
points for 9 ML GdSt LB
film and solid line is the curve calculated with an electron density
profile (shown in the inset) assuming two stearic acid tails attached
to a gadolinium ion. The dashed line is the calculated reflectivity
curve with the expected electron density profile corresponding to a model 
where three stearic acid tails attached to a gadolinium ion.
Reflectivity data of a 51 ML sample (line+circle) is also shown for
comparison. This data has been shifted down for clarity (refer text).}

\caption {Magnetization curves as a function of (H/T) for the (a)
out-of-plane direction measured at 5K (open circle), 10K
(solid circle) and 20K (star) with fit using Eq. (1) (solid line),
and for the (b) in plane direction measured at 2K (solid triangle),
5K (open circle),
10K (solid circle) and 20K (star). In-plane data shows the absence of
scaling observed in (a). The inset shows log-linear plot of the
in-plane saturation magnetization (M$_s$) expressed in Bohr
magnetron per gadolinium ion at various temperatures and solid line
is the linear fit (refer text).}

\caption {In-plane magnetization as a function of applied field for
101ML GdSt LB film measured at temperatures 2K (solid triangle), 5K (open
circle), 10K (solid circle) and 20K (star). Solid line is the
corresponding fit with the modified Brillouin function as given
in Eq. (3). The dashed line is the best-fitted Brillouin
function with field independent exchange (refer text).}

\end {figure}

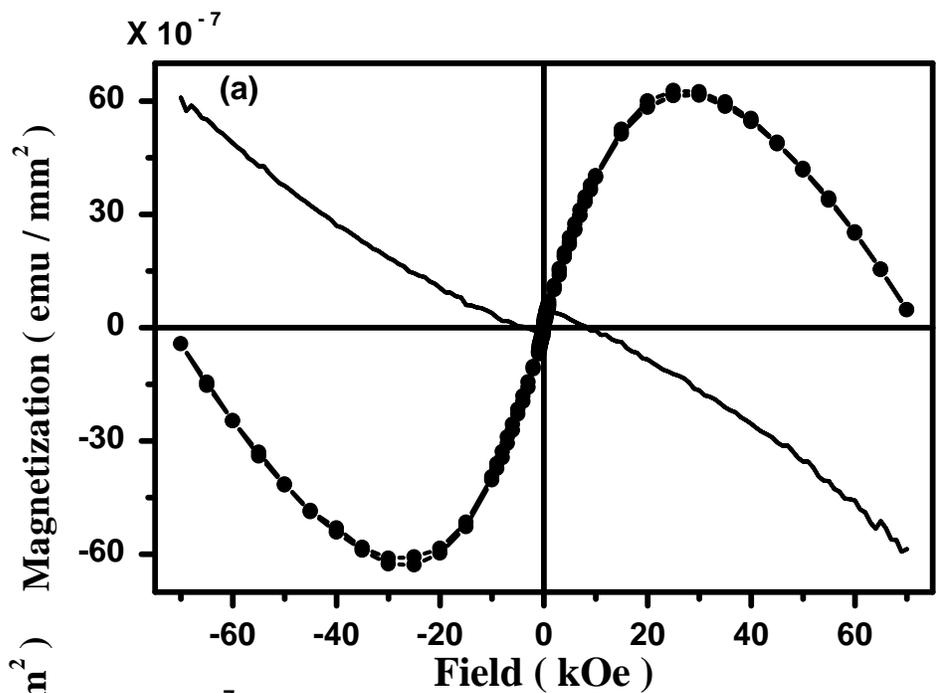
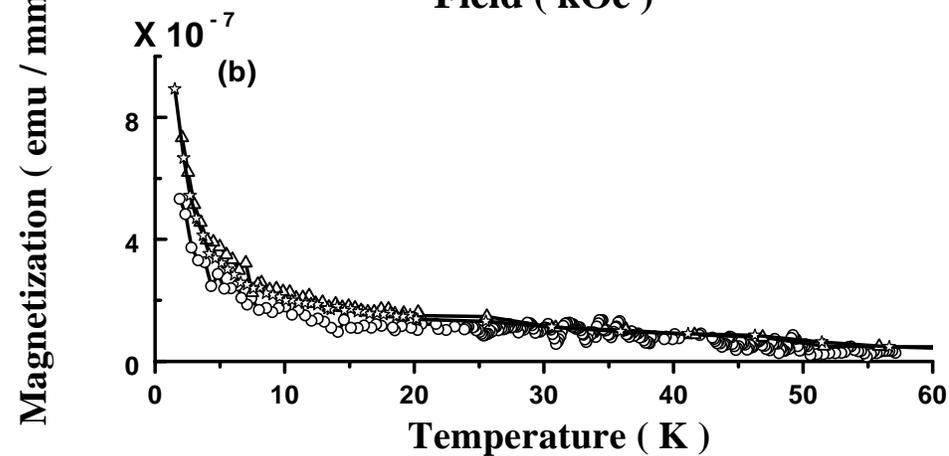
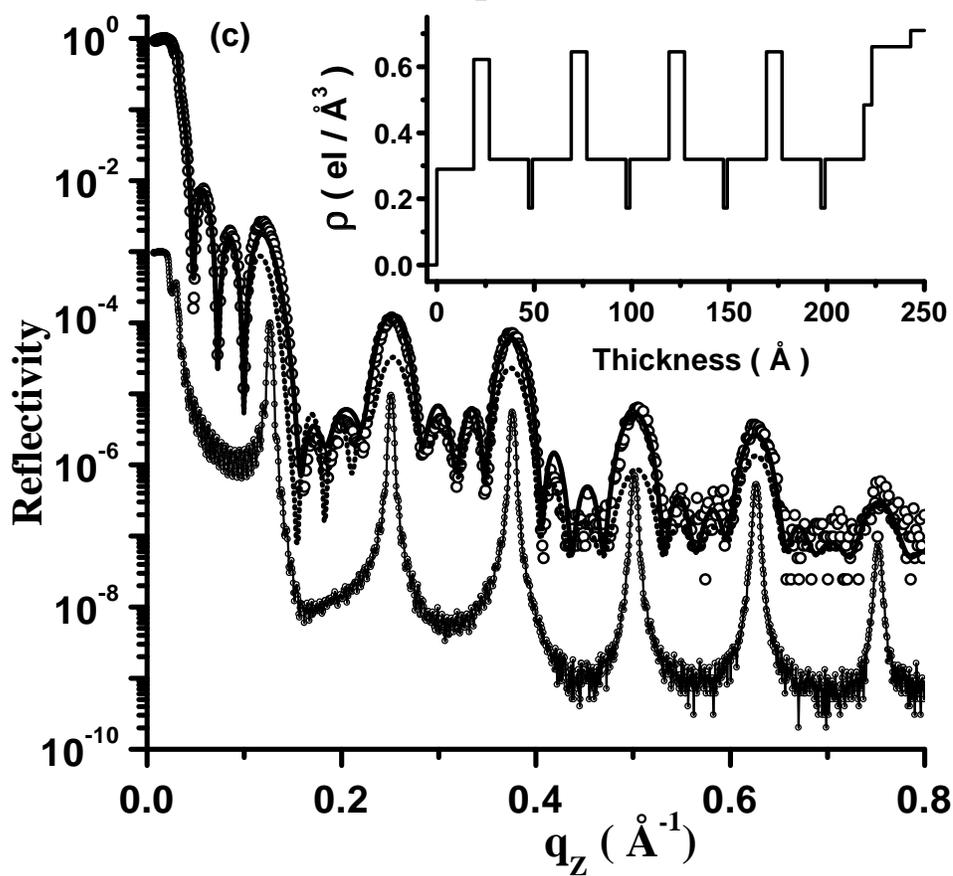
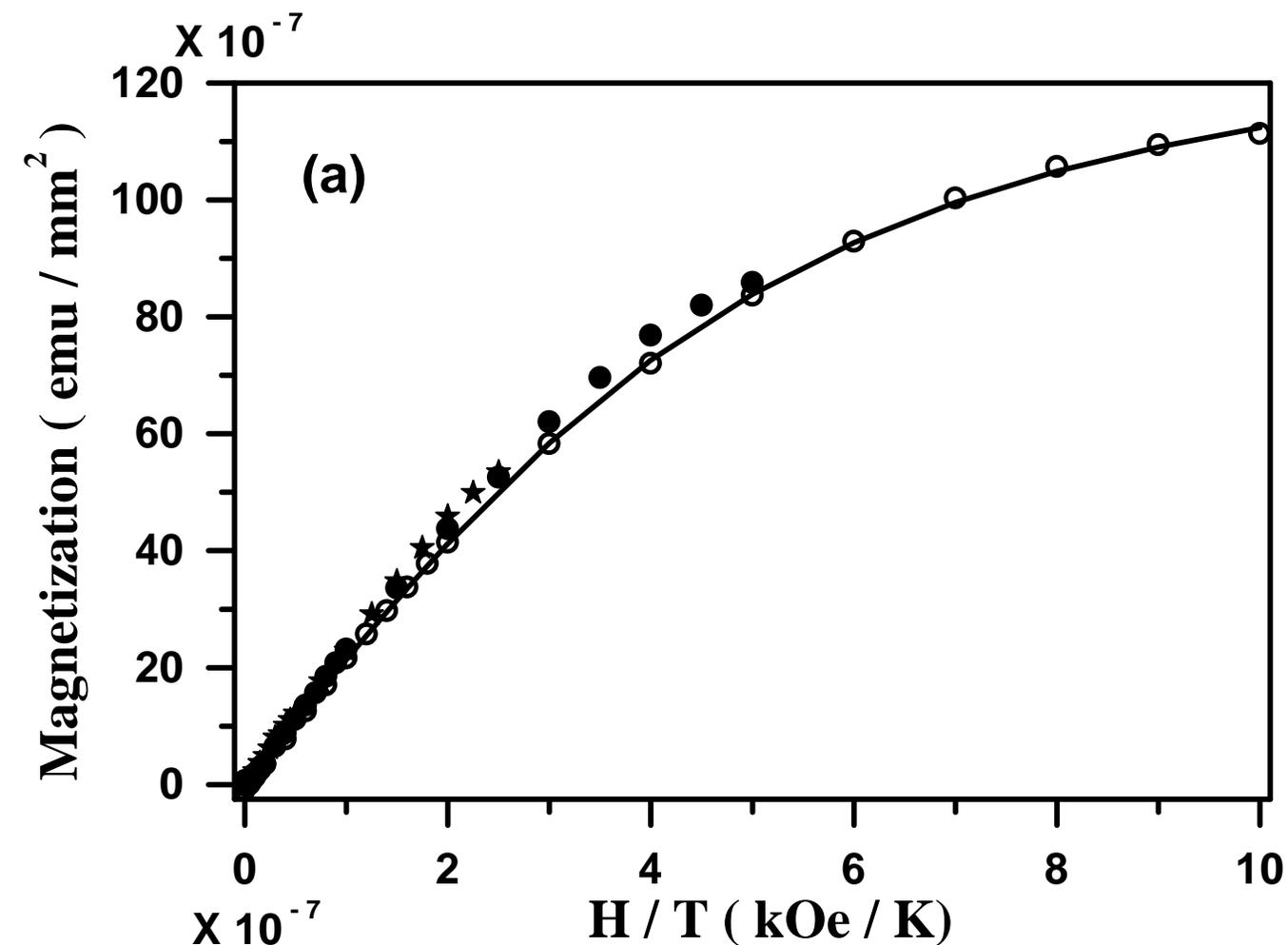
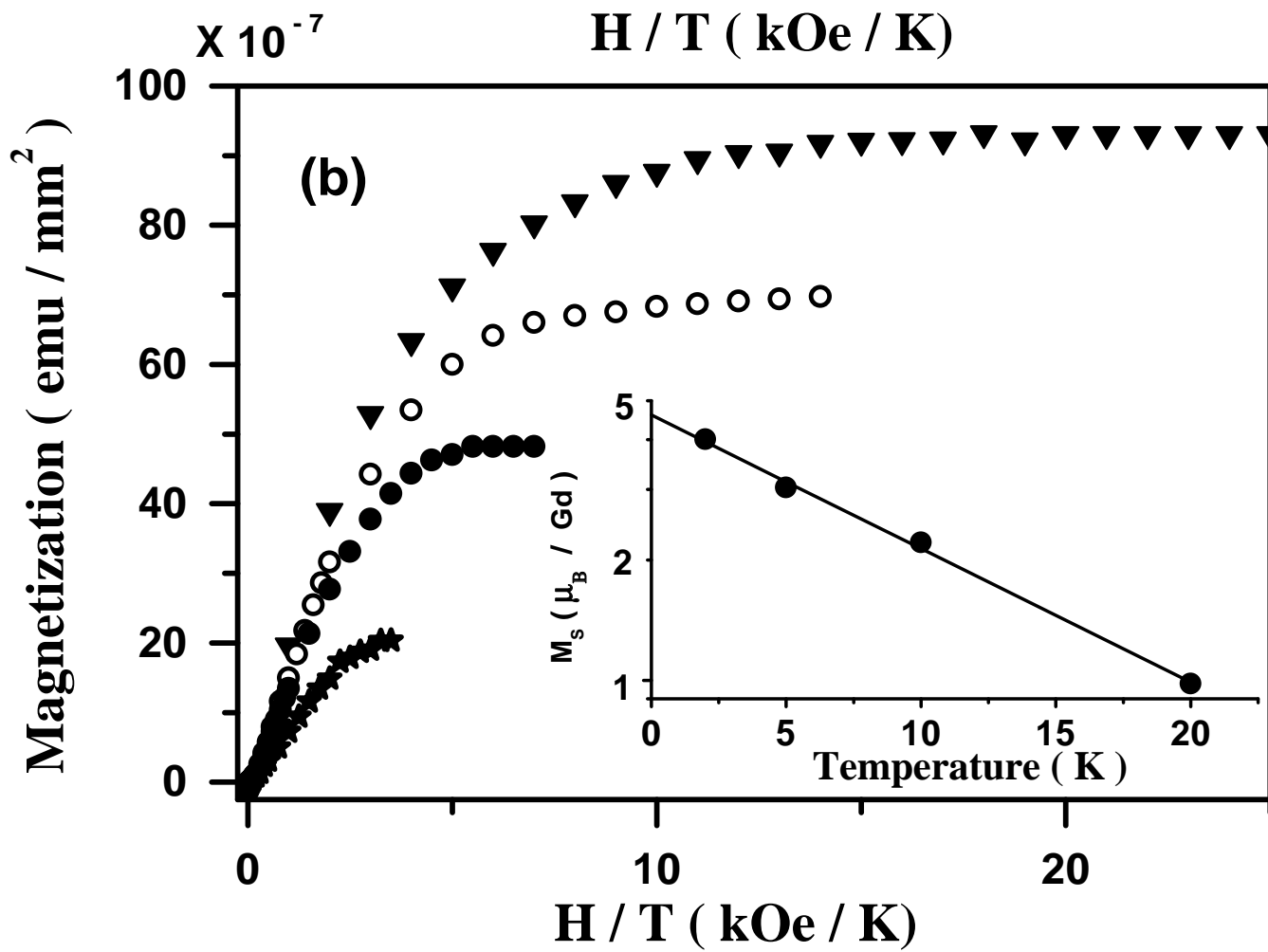
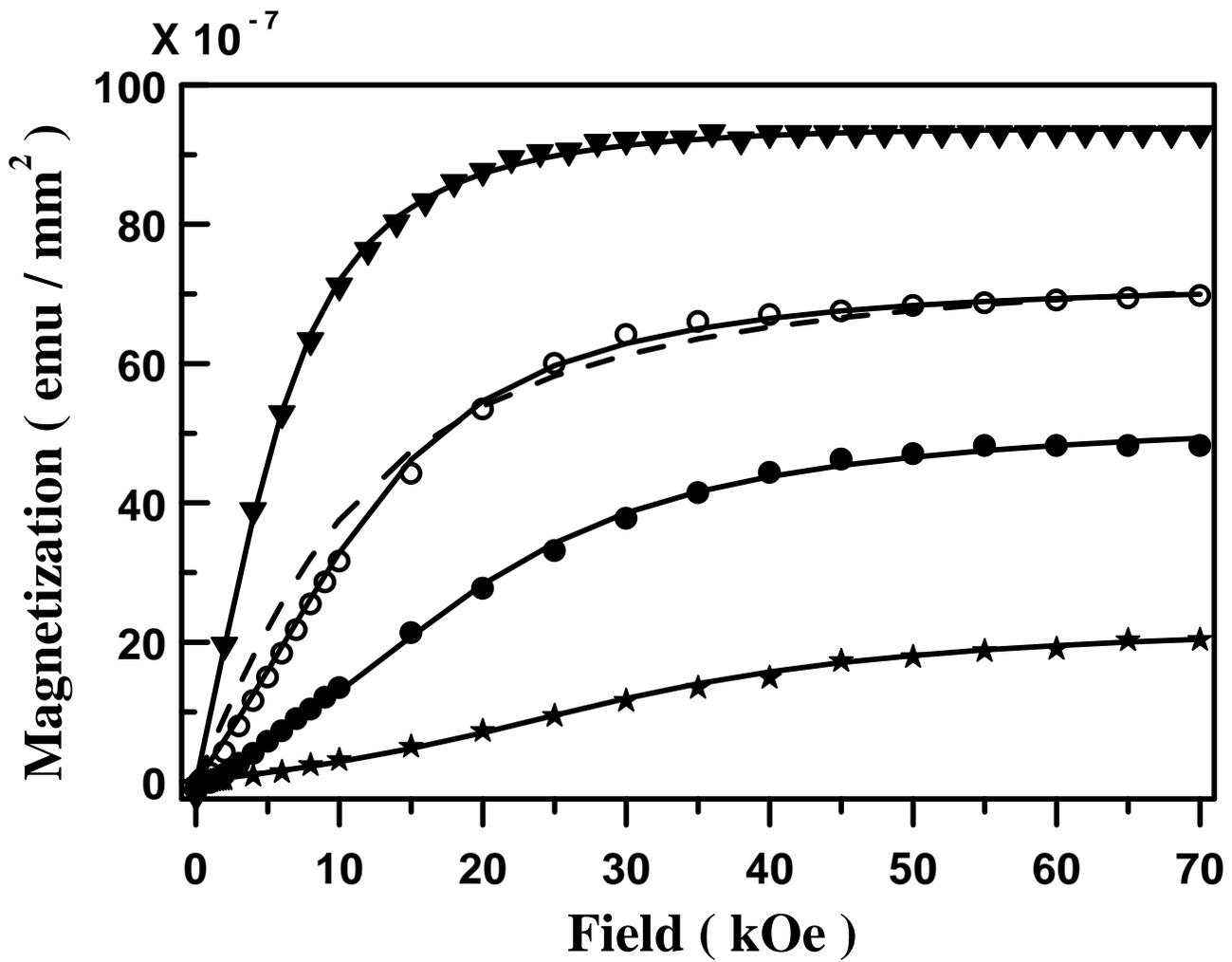
\end{document}